\documentclass[a4paper,11pt]{article}
\usepackage{jcappub} 
\usepackage{lineno}
\usepackage{tikz}
\usepackage{amsmath}
\usepackage{xcolor}
\usepackage[normalem]{ulem}
\usepackage{fontawesome}

\arxivnumber{2508.20606}
\title{\boldmath Incorporating curved geometry in cosmological simulations}

\author{Julian Adamek}
\author{and Renan Boschetti}
\affiliation{Institut f\"ur Astrophysik, Universit\"at Z\"urich\\Winterthurerstrasse 190, 8057 Z\"urich, Switzerland}

\emailAdd{julian.adamek@uzh.ch}
\emailAdd{renan.boschetti@gmail.com}

\abstract{Spatial curvature is one of the fundamental cosmological parameters that is routinely constrained from observations. The forward modelling of observations, in particular of large-scale structure, often relies on large cosmological simulations. While the so-called separate universe approach allows one to account for the effect of curvature on the expansion rate in small sub-volumes, the non-Euclidean geometry is harder to accommodate. It becomes important when observables are computed over large distances, e.g.\ when photons travel to us from high redshift. Here we present a fully relativistic framework to run cosmological simulations for curved spatial geometry. The issue of consistent boundary conditions is solved by embedding a spherical cap of the curved spacetime into a hole within a flat exterior, where it can undergo free expansion. The geometric nature of gravity is made explicit in our framework, allowing for a consistent forward modelling of observables inside the curved patch. Our methodology would also work with any Newtonian code to a good approximation, requiring changes only to the initial conditions and post-processing.}

\begin{document}
\maketitle
\flushbottom

\section{Introduction}
\label{sec:intro}

Spatial flatness is a critical prediction of inflation \cite{Guth:1980zm} that should be tested and not assumed from the outset. Observations of anisotropies in the cosmic microwave background (see, e.g.\ \cite{Boomerang:2000jdg} for early results and \cite{Planck:2018vyg,ACT:2025fju} for recent data) put strong constraints on curvature which are, however, somewhat model dependent. Some analyses even suggest a preference for spatial curvature; see \cite{DiValentino:2019qzk} for an example. Largely independent constraints can be obtained from distance measurements in the late Universe and certain probes of large-scale structure. For example, recent measurements of baryon acoustic oscillations (BAO) by the Dark Energy Spectroscopic Instrument (DESI) constrain the curvature parameter, $\Omega_K$, to $0.025\pm0.041$ at $68\%$ confidence \cite{DESI:2025zgx}. It has been argued that even a curvature as small as $0.2\%$, corresponding to a radius of curvature of $21 H_0^{-1}$, would be important for cosmological constraints on the neutrino mass \cite{Chen:2025mlf}. A conclusive detection of spatial curvature would have profound implications for the origin of our Universe of course, shattering the prevailing paradigm of inflation and offering a rare glimpse at the conditions from which it emerged. 

With the wealth of observational data from surveys such as DESI, Euclid \cite{EUCLID:2011zbd,Euclid:2024yrr}, or the Vera C.\ Rubin observatory \cite{LSSTScience:2009jmu}, obtaining direct constraints that are robust below the percent level will require careful modelling. In the context of large-scale structure, the forward modelling typically relies on sophisticated simulations of the gravitational dynamics and the kinematical projection effects on the past light cone of our observations. Such simulations are employed, for example, to estimate covariances \cite{Norberg:2008tg}, to train emulators that can be used for parameter inference \cite{Jamieson:2022lqc,Euclid:2020rfv}, or for simulation-based inference \cite{Bairagi:2025sux,Tucci:2023bag}.

In the late Universe, curvature has two main effects on large-scale structure probes. First, it affects the expansion history and therefore the growth rate. This effect is relatively straightforward and has been studied with the so-called separate universe approach \cite{Li:2014sga,Wagner:2014aka}. Since matter only moves a few megaparsecs in one Hubble time, distances can be considered approximately Euclidean on the scales of interest for structure formation. Simulations of small Euclidean sub-volumes that account for the correct expansion history on a curved background therefore capture the evolution of matter well.

The second effect concerns the non-Euclidean distance and volume when considering large scales. These are the scales probed by the photons as they travel cosmological distances. Intuitively, given a standard ruler such as the BAO scale, a curved geometry implies that the circumference of a circle around the observer is either larger (for an open Universe) or smaller (for a closed Universe) than $\pi$ times the geometric diameter when measured with the same ruler. Similar arguments hold for the surface of a sphere or for the physical volume of a redshift bin. This poses the question of how the sub-volumes comprising the separate universe simulations could be stitched together consistently to build the full light cone of a curved Universe.

Further interesting aspects of curvature concern the early Universe. Since standard inflation would always lead to a flat spatial geometry, it cannot be invoked to predict the primordial power spectrum, and thus we need to modify our prescription for setting initial conditions. Even if some inflation occurred, perturbation modes close to the curvature scale would always remain sensitive to pre-inflationary physics. In addition, the basis of mode functions that describe the perturbations also depends on the spatial geometry, with the spectrum becoming discrete in a Universe that is spatially closed. These aspects become important at exactly the scales we are interested in, and using them to gain insights into the origin of the Universe is certainly very compelling. However, from the point of view of simulations, they are relevant only for setting initial conditions of matter perturbations, which is not the main focus of this work.

In most simulation frameworks, in particular $N$-body simulations, accounting for a curved spatial geometry is not straightforward. The reason is that boundary conditions need to be imposed, and almost all frameworks have settled on periodic boundaries. Beyond the scale of periodicity, the volume therefore necessarily behaves like in a flat geometry. Although changing boundary conditions could be done in principle (see, e.g.\ \cite{Racz:2018nrx} for an interesting example), it is often not desired because too much of the established code infrastructure depends on this assumption. We will therefore work within the constraints of periodic boundaries, meaning that we must incorporate curvature somehow below the scale of periodicity.

The remainder of this paper is structured as follows. In Sec.\ \ref{sec:method} we present our methodology, which is essentially based on translating a specific Lema\^itre--Tolman--Bondi model into the coordinates used in simulations. In Sec.\ \ref{sec:results} we show some first numerical results to validate our framework. We conclude in Sec.\ \ref{sec:conclusion}. Some details of our calculations are relegated to App.\ \ref{app:2nd-order}. As a side product of our work, we present in App.\ \ref{app:Schwarzschild} the Schwarzschild metric in a new coordinate system motivated by our setup.

\section{Methodology}
\label{sec:method}

The basic idea of our approach is to embed a patch of a curved Friedmann--Lema\^itre--Robertson--Walker (FLRW) spacetime, which shall contain any observations we wish to model, into an exterior solution that is spatially flat, so that periodic boundaries can be employed. The junction conditions for joining different solutions of Einstein's equations have been worked out in full generality in a work by Israel \cite{Israel:1966rt}. Using these prescriptions, one finds that any spherical patch of a FLRW spacetime with pressureless matter and arbitrary curvature, truncated at some fixed comoving radius, can be embedded into the Schwarzschild solution. The resulting spacetime is an exact solution to Einstein's equation describing a uniform ball of matter undergoing free expansion in vacuum.

\begin{figure}[t!]
\centering
\includegraphics[width=0.45\textwidth]{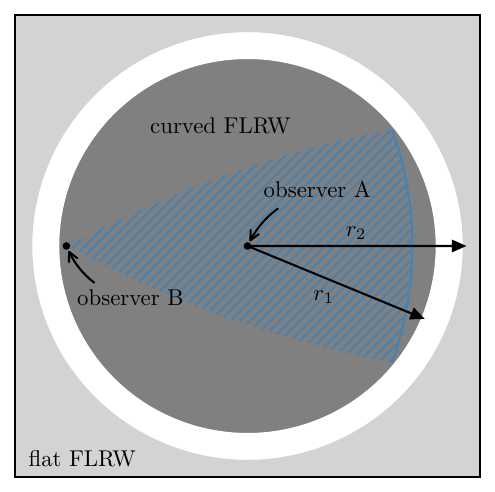}
\caption{We embed a sphere of radius $r_{1}$ and initial overdensity $\delta_{1}$ inside a ``hole'' of radius $r_{2}$ carved from a flat FLRW exterior. This setup allows periodic boundary conditions in a cubic simulation box. Observer A at the center sees a curved FLRW spacetime on the full sky out to the distance $r_1$. Observer B close to the perimeter sees the same curved FLRW spacetime when looking inward. Observations in the curved region can reach higher redshift in this case, but only for a field of view smaller than a quarter of the full sky (example illustrated as hatched region).}
\label{fig:setup}
\end{figure}

This solution is not yet what we want since Schwarzschild only becomes flat asymptotically at spatial infinity. However, we can repeat the process once more and embed the Schwarzschild solution (which embeds the curved FLRW model) into an outer flat FLRW model. We can then fit the entire system into a cubic simulation box with periodic boundary conditions. The setup is illustrated in Fig. \ref{fig:setup}. A similar idea was originally proposed by Einstein \& Straus \cite{Einstein:1945id} to model an expanding universe filled with stars, and we emphasise once again that this construction is exact for dust-like matter. For the second matching radius to be larger than the radius of the curved interior FLRW patch, the curvature of the latter has to be positive, modelling a closed Universe. For small values of the curvature parameter, the two matching radii approach the same value, merging into one as the curvature goes to zero and the solution simply becomes a continuous flat FLRW spacetime. Open FLRW patches can also be embedded, but one needs to add some compensating mass shell inside the vacuum region to obtain a plausible sequence of matching radii. In this work, we avoid this minor complication and focus on embedding closed FLRW patches, but we keep our derivations general in most places so that the open case can be treated with minimal adjustments.

In Fig.~\ref{fig:setup}, we also indicate two possible choices to place an observer inside the curved patch. Observer A is at the center of the patch, observing a curved Universe in all directions up to the inner matching radius $r_1$. While this might be the most obvious choice, we have to keep in mind that simulations always need to trade volume against resolution (for a fixed number of particles). If the observations we want to model go to some maximum redshift $z_\mathrm{max}$, this determines the minimum value of $r_1$ and thus puts a corresponding limit on the resolution. However, if the observations cover only a part of the sky, we may place the observer off centre to allow a larger redshift coverage or, equivalently, a smaller $r_1$ for the same $z_\mathrm{max}$. This situation is illustrated with observer B in the figure, who sits close to the perimeter of the curved patch and observes a cone-shaped field of view towards the far side of the patch. 

The remainder of this section deals with the question of how to set up this system precisely. We start by noting that for pressureless matter, often called pure dust, the entire spacetime solution is known analytically, and the metric in matter-comoving synchronous coordinates is given by the Lema\^itre--Tolman--Bondi (LTB) metric (see \cite{plebanski2024introduction} for details). We shall set initial conditions deep in the matter era, so this solution is a valid starting point. However, simulations are not performed in comoving synchronous coordinates, so we need to work out a suitable coordinate transformation. We choose to work in the Poisson gauge adapted to the exterior flat FLRW model, where the metric takes the form
\begin{equation}
\label{eq:metric}
    ds^2 = -e^{2\psi} dt^2 + a^2(t) e^{-2\phi} \left[dr^2 + r^2 d\Omega^2\right]\,,
\end{equation}
where $\psi$ and $\phi$ are the two Bardeen potentials and $a(t)$ is the scale factor of the exterior FLRW that, from the point of view of the simulation code, is the relevant cosmology. We use units where $c = 1$. For most of the discussion presented here we do not consider matter perturbations inside the curved domain yet, although we shall get to that at the end. For simplicity, we therefore ignore vector and tensor perturbations and initially consider $\phi$ and $\psi$ as functions of $t$ and $r$ only.

To understand how spatial curvature is incorporated in this picture, it is instructive to consider the Hamiltonian constraint. In any foliation of spacetime, it takes the form ${}^3\mathcal{R} - K_{ij} K^{ij} + K^2 = 16 \pi G \rho$, where ${}^3\mathcal{R}$ is the curvature of the three-dimensional constant-time hypersurface, $K_{ij}$ is the extrinsic curvature tensor describing the embedding into the four-dimensional manifold, $K$ is its trace, and $\rho$ is the matter density measured along the timelike hypersurface normal (see, e.g.\ \cite{BaumgarteShapiro} for details). For the Poisson gauge, the Hamiltonian constraint becomes
\begin{equation}
\label{eq:Hamiltonianconstraint}
    4 \nabla^2 \phi + 2 \left(\nabla\phi\right)^2 + 6 \left(\partial_t \phi - H\right)^2 e^{-2 \psi} = 16 \pi G \rho\,,
\end{equation}
where $H = \partial_t \ln a$ and $\nabla$ is the covariant derivative on the three-dimensional hypersurface, i.e.\ $\nabla^2 \phi = a^{-2} e^{2\phi} \left[\partial_r^2\phi +\frac{2}{r} \partial_r \phi - \left(\partial_r \phi\right)^2\right]$ and $\left(\nabla\phi\right)^2 = a^{-2} e^{2\phi} \left(\partial_r\phi\right)^2$. The first two terms of Eq.~\eqref{eq:Hamiltonianconstraint} represent the three-curvature ${}^3\mathcal{R}$, whereas the third term comes from the extrinsic curvature. In the flat exterior FLRW solution, the two potentials are set to zero and the Hamiltonian constraint reduces to the Friedmann equation. In the curved patch, the potentials are non-zero and $\phi$ gives rise to spatial curvature. As a minor point, note that spacetime is sliced in the rest frame of the exterior FLRW and hence the constant-time hypersurfaces are slightly tilted with respect to the homogeneous slices of the closed patch. This subtlety leads to a slightly non-uniform curvature on constant-time slices (beyond linear order), as well as to some gravitational time delays, which we will have to deal with later.

\subsection{The Lema\^itre--Tolman--Bondi solution}

The LTB metric,
\begin{equation}
\label{eq:LTB}
    ds^2 = -dt^2 + \frac{\left(\partial_r R\right)^2}{1 + 2 E(r)} dr^2 + R^2 d\Omega^2\,,
\end{equation}
describes an isotropic, but not homogeneous, spacetime in comoving synchronous coordinates for a pressureless matter source. Einstein's equations yield the parametric solutions
\begin{eqnarray}
    R(t,r) &=& -\frac{G M(r)}{2E(r)} \left(1-\cos\eta\right)\,,\\
    \left(\eta - \sin\eta\right)^{2/3} &=& -\frac{2E(r)}{G^{2/3}M^{2/3}(r)} t^{2/3}\,,
\end{eqnarray}
for the two functions $E(r)$ and $R(t, r)$. Following \cite{Adamek:2015hqa}, we set the mass profile to
\begin{equation}
    G M(r) = \frac{3}{2} H^2 a^3 \int_0^r \tilde{r}^2 \left[1 + \delta(\tilde{r})\right]d\tilde{r}\,,
\end{equation}
where
\begin{equation}
    \delta(r) = \begin{cases} 
          \delta_1 & r < r_1 = r_2 \left(1 + \delta_1\right)^{-1/3}\\
          -1 & r_1 < r < r_2 \\
          0 & r > r_2\,.
       \end{cases}
\end{equation}
Here, $\delta_1$ is a uniform initial density contrast inside the region $r < r_1$ whose relation with curvature is made explicit later.
We therefore have $G M(r) = \frac{1}{2} H^2 a^3 r^3 \left[1 + f(r)\right]$ with
\begin{equation}
    f(r) = \begin{cases} 
          \delta_1 & r < r_1 \\
          \frac{r_2^3}{r^3} - 1 & r_1 < r < r_2 \\
          0 & r > r_2\,.
       \end{cases}
\end{equation}
Note that the scale factor $a$ and the associated Hubble function $H$ still refer to the flat outer FLRW region, where the synchronous and Poisson gauges coincide.

At first order in perturbation theory, the gauge transformation between the comoving synchronous gauge and the Poisson gauge is well known and has been applied to the same problem in \cite{Adamek:2015hqa}. Here we wish to extend this calculation to second order. The reason for this is two-fold. First, it allows us to set initial conditions more accurately, making our method more robust when curvature is large. Second, we are ultimately interested in the coupling between perturbations and curvature. If we count curvature itself as one order in perturbation theory, then this coupling appears at second order, and is therefore potentially as important as second-order corrections from curvature itself.

Proceeding as in \cite{Adamek:2015hqa}, we rescale the coordinate $r$ so that at some initial time $t_\mathrm{in}$ we have $R(t_\mathrm{in}, r) = a(t_\mathrm{in}) r \equiv a_\mathrm{in} r$.
Eliminating $\eta$ perturbatively to second order in $f(r)$, the function $E(r)$ that appears in the LTB metric then reads
\begin{equation}\label{eq:E2nd}
    E(r) = -\frac{5}{6} H_\mathrm{in}^2 a_\mathrm{in}^2 f(r) \left[1 - \frac{1}{7} f(r)\right] r^2\,,
\end{equation}
The corresponding solution for $R(t,r)$ reads
\begin{equation}\label{eq:R2nd}
    R(t,r) = a(t) \left[1 + \frac{1}{3} \left(1 - \frac{a(t)}{a_\mathrm{in}}\right) f(r) \left(1 - \left(\frac{1}{3}-\frac{1}{7}\frac{a(t)}{a_\mathrm{in}}\right)f(r)\right)\right] r\,.
\end{equation}
Given the form of the LTB metric, Eq.~\eqref{eq:LTB}, and observing the fact that $f(r)$ is constant inside $r_1$, we recover the metric of a curved FLRW spacetime for $r < r_1$ and can immediately read off the curvature parameter $K = \frac{5}{3} H_\mathrm{in}^2 a_\mathrm{in}^2 \delta_1 \left(1 - \frac{1}{7}\delta_1\right)$. On the initial hypersurface, the curved Friedmann solution has a Hubble rate $\tilde{H}$ that satisfies
\begin{equation}
\label{eq:curvedFriedmann}
    \tilde{H}^2 + \frac{K}{a_\mathrm{in}^2} = H_\mathrm{in}^2 \left(1 + \delta_1\right)\,,
\end{equation}
and hence $\tilde{\Omega}_\mathrm{m} \simeq 1 + \frac{5}{3}\delta_1\left(1 + \frac{11}{21}\delta_1\right)$, $\tilde{\Omega}_K \simeq -\frac{5}{3} \delta_1 \left(1 + \frac{11}{21}\delta_1\right)$. Here and in the following, we use a tilde to indicate cosmological parameters in the curved patch. The expansion rate is somewhat reduced at $\tilde{H} \simeq H_\mathrm{in} \left(1 - \frac{1}{3}\delta_1 + \frac{4}{63} \delta_1^2\right)$.

\subsection{Coordinate transformation}

Up to second order in perturbation theory, the relation between synchronous gauge and Poisson gauge is written as
\begin{eqnarray}
    t^\mathrm{(syn)} &=& t^\mathrm{(P)} + T_1(t^\mathrm{(P)}, r^\mathrm{(P)}) + T_2(t^\mathrm{(P)}, r^\mathrm{(P)})\,,\label{eq:t-coordtra}\\
    r^\mathrm{(syn)} &=& r^\mathrm{(P)} + L_1(t^\mathrm{(P)}, r^\mathrm{(P)}) + L_2(t^\mathrm{(P)}, r^\mathrm{(P)})\,.\label{eq:r-coordtra}
\end{eqnarray}
To avoid clutter, labels (syn) and (P) will only be used when necessary, that is, when two different coordinate types appear in the same equation. Explicit expressions for the coordinate transformations $T_1$, $T_2$, $L_1$, $L_2$, and the resulting expressions for $\phi$ and $\psi$ are derived in App.~\ref{app:2nd-order}. To obtain the initial density, we can either work out the gauge transformation of the stress-energy tensor or simply evaluate the Hamiltonian constraint~\eqref{eq:Hamiltonianconstraint} given the potentials. The peculiar velocity can be easily inferred by noting that
\begin{equation}
    u_\mu^\text{(P)} = u_\nu^\text{(syn)} \frac{dx^\nu_\text{(syn)}}{dx^\mu_\text{(P)}}\,,
\end{equation}
with $u_\nu^\text{(syn)} = -(1, 0, 0, 0)$. Our simulation code uses the canonical momentum as the phase-space coordinate for the particles, given by $q_i = m u_i^\text{(P)}$.

\subsection{Cosmological parameters in the simulation}

With the relation between the Bardeen potentials and the curved background cosmology in hands, we can initialise the simulation by providing the cosmological parameters and initial redshift as seen by the observer in the curved Universe. These need to be translated into a corresponding set of parameters for the exterior flat FLRW spacetime, as this is the one that is evolved from the point of view of the simulation code. The first step is to obtain the cosmological parameters at the initial time where the coordinate transformation is applied. Working in the curved patch, we have
\begin{eqnarray}
    \tilde{\Omega}_K &=& \frac{\tilde{\Omega}_K^0 \left(1+\tilde{z}_\mathrm{in}\right)^2}{\tilde{\Omega}_\mathrm{m}^0 \left(1+\tilde{z}_\mathrm{in}\right)^3 + \tilde{\Omega}_K^0 \left(1+\tilde{z}_\mathrm{in}\right)^2 + \tilde{\Omega}_\Lambda^0}\,,\\
    \tilde{\Omega}_\mathrm{m} &=& \frac{\tilde{\Omega}_\mathrm{m}^0 \left(1+\tilde{z}_\mathrm{in}\right)^3}{\tilde{\Omega}_\mathrm{m}^0 \left(1+\tilde{z}_\mathrm{in}\right)^3 + \tilde{\Omega}_K^0 \left(1+\tilde{z}_\mathrm{in}\right)^2 + \tilde{\Omega}_\Lambda^0}\,,\\
    \tilde{\Omega}_\Lambda &=& \frac{\tilde{\Omega}_\Lambda}{\tilde{\Omega}_\mathrm{m}^0 \left(1+\tilde{z}_\mathrm{in}\right)^3 + \tilde{\Omega}_K^0 \left(1+\tilde{z}_\mathrm{in}\right)^2 + \tilde{\Omega}_\Lambda^0}\,,\\
    \tilde{H} &=& \tilde{H}_0 \sqrt{\tilde{\Omega}_\mathrm{m}^0 \left(1+\tilde{z}_\mathrm{in}\right)^3 + \tilde{\Omega}_K^0 \left(1+\tilde{z}_\mathrm{in}\right)^2 + \tilde{\Omega}_\Lambda^0}\,,
\end{eqnarray}
where the label $0$ indicates parameters evaluated at present time and we assume that the density of radiation can be neglected. Furthermore, we choose $\tilde{z}_\mathrm{in}$ so that $\tilde{\Omega}_\Lambda$ is sufficiently small and we can use the pure dust solution to compute the coordinate transformations on the initial hypersurface. We also note that the proper look-back time in matter-synchronous time is given by
\begin{equation}
\label{eq:lookbacktime}
    t^\text{(syn)}_0 - t^\text{(syn)}_\mathrm{in} = \tilde{H}_0^{-1} \int_0^{\tilde{z}_\mathrm{in}} \frac{dz'}{1+z'} \left[\tilde{\Omega}_\mathrm{m}^0 (1+z')^3 + \tilde{\Omega}_K^0 (1+z')^2 + \tilde{\Omega}_\Lambda^0\right]^{-1/2}\,.
\end{equation}

We can now fix the initial cosmological parameters of the flat model. First, we determine the auxiliary variable $\delta_1$ by inverting the relation with $\tilde{\Omega}_K$, finding $\delta_1 \simeq -\frac{3}{5} \tilde{\Omega}_K \left(1 + \frac{11}{35}\tilde{\Omega}_K\right)$. Using Eq.~\eqref{eq:curvedFriedmann} we then have
\begin{equation}
    H_\mathrm{in} \simeq \tilde{H} \left(1 + \frac{1}{3}\delta_1 + \frac{1}{21}\delta_1^2\right) \simeq \tilde{H}\left(1 - \frac{1}{5}\tilde{\Omega}_K-\frac{8}{175}\tilde{\Omega}_K^2\right)\,.
\end{equation}
Relating cosmological parameters to proper densities we have $\tilde{H}^2 \tilde{\Omega}_\mathrm{m} = H_\mathrm{in}^2 \Omega_\mathrm{m}\left(1+\delta_1\right)$ and $\tilde{H}^2 \tilde{\Omega}_\Lambda = H_\mathrm{in}^2 \Omega_\Lambda$, accounting for the fact that the matter density in the exterior is lower but the vacuum energy density is the same everywhere. Therefore,
\begin{eqnarray}
    \Omega_\mathrm{m} &\simeq& \tilde{\Omega}_\mathrm{m} \left(1 - \frac{5}{3} \delta_1 + \frac{40}{21} \delta_1^2\right) \,\simeq\, \tilde{\Omega}_\mathrm{m} \left(1 + \tilde{\Omega}_K + \tilde{\Omega}_K^2\right)\,,\\
    \Omega_\Lambda &\simeq& \tilde{\Omega}_\Lambda \left(1 - \frac{2}{3} \delta_1 + \frac{5}{21} \delta_1\right) \,\simeq\, \tilde{\Omega}_\Lambda \left(1 + \frac{2}{5} \tilde{\Omega}_K + \frac{37}{175} \tilde{\Omega}_K^2\right)\,.
\end{eqnarray}

The last thing we have to do is to scale all parameters back to today's values in the flat exterior spacetime. Or, more precisely, we need to tell the code for how long (in terms of exterior coordinate time) the simulation needs to run until the observer reaches $\tilde{z} = 0$. This point turns out to be a bit subtle, because the exterior coordinate time is not synchronised with the proper time in the curved patch. Adding to the complication, the matter-synchronous clock runs differently in different places when viewed in the coordinates of the Poisson gauge.

The latter issue is fixed by choosing the location of the observer within the patch. Consider first an observer at the center (indicated as observer A in Fig.~\ref{fig:setup}). This observer remains at rest in both coordinate systems used here, and the relation between matter-synchronous time (eigentime) and coordinate time is given by the lapse function evaluated at $r = 0$. Specifically, the matter-synchronous clock runs slower, with $dt^\text{(syn)} = e^{\psi} dt^\text{(P)}$. This can also be seen from the generators $T_1$, $T_2$ of the time coordinate transformation.

\begin{figure}[t!]
\centering
\includegraphics[width=0.45\textwidth]{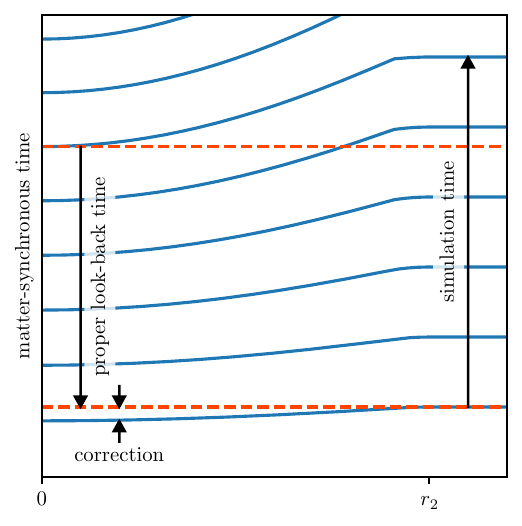}
\caption{Illustration of the relation between the time slicing in matter-synchronous coordinates (horizontal lines correspond to constant synchronous time) and the one in Poisson gauge (blue lines), as a function of spacetime position. The x-axis indicates comoving distance from the origin. For distances larger than the outer matching radius, $r_2$, both time coordinates coincide exactly. Inside the curved patch, the matter-synchronous clocks run slower than the coordinate time of Poisson gauge. To compute the simulation time, one first chooses the proper look-back time to initial conditions for the observer, placed at the origin in this example. Then one adds a small correction due to the fact that the initial constant-time hypersurface in Poisson gauge is slightly tilted. Finally, one accounts for the time dilation between exterior coordinate clock and matter clock at the observer position, see text for details. We make some extreme parameter choices to exaggerate the effects; for realistic values used in simulations the differences in time slicing would not be visible to the eye.}
\label{fig:timeslicings}
\end{figure}

The situation is illustrated in Fig.~\ref{fig:timeslicings}, showing the constant-time slices of the Poisson gauge in matter-synchronous coordinates. Choosing an initial redshift in the curved patch corresponds to setting the proper look-back time to the initial conditions. Since the constant-time slices are slightly tilted between the two coordinate frames, the simulation is initialised on a slightly different hypersurface. We choose the clocks to coincide in the exterior flat FLRW where the two coordinate systems match identically. This means that at the center of the curved patch, the proper look-back time to the initial hypersurface of the simulation needs to be corrected by a small amount that is given by the time coordinate transformation we have already computed. Finally, the simulation time is computed by taking into account the integrated effect of time dilation,
\begin{equation}
\label{eq:t0}
    t^\text{(syn)}_0 - t^\text{(syn)}_\mathrm{in} \underbrace{- T_1(t^\text{(syn)}_\mathrm{in}, 0) - T_2(t^\text{(syn)}_\mathrm{in}, 0)}_{\text{correction term}} = \int_{t^\text{(P)}_\mathrm{in} = t^\text{(syn)}_\mathrm{in}}^{t^\text{(P)}_0} e^\psi dt^\text{(P)}\,.
\end{equation}
In this equation, we solve for the unknown upper bound $t_0^\text{(P)}$ of the integral, which can be done with a simple root finding algorithm. One might also try to use the time coordinate transformation at the final time, but this is less robust for cases with large curvature. The expressions we derived for the coordinate transformation are valid in a small region around the initial time.

To evaluate the lapse perturbation $\psi$ in Eq.~\eqref{eq:t0}, we use the expressions derived in App.~\ref{app:2nd-order}. In the presence of a cosmological constant, we also account for the modified linear growth of the first-order solution in the exterior flat $\Lambda$ cold dark matter (CDM) model. If the exact behaviour of $\psi$ is difficult to estimate, one can also simply run a small low-resolution simulation and evaluate the integral on the right-hand side of Eq.~\eqref{eq:t0} on the fly.

For an observer close to the perimeter of the curved patch, such as observer B in Fig.~\ref{fig:setup}, the difference between the coordinate time in the Poisson gauge and eigentime becomes very small. In a closed Universe, it therefore reaches redshift $\tilde{z} = 0$ sooner (in terms of simulation time) than the observer at the center.

Once the timeline of the simulation is fixed, we can infer relevant simulation parameters such as $\Omega_\mathrm{m}^0$, $\Omega_\Lambda^0$, $H_0$, $z_\mathrm{in}$, and so on, by evolving the exterior flat FLRW cosmology from the initial time.

\subsection{Particle initial data and relativistic mass defect}

In $N$-body simulations, the matter density is represented through a particle ensemble. For performance considerations, one typically chooses identical masses for all particles. To set initial conditions, the most common procedure is to lay down a homogeneous particle distribution, e.g.\ a regular crystal-like arrangement. Then the particle positions are perturbed to obtain the desired initial density field; see \cite{Angulo:2021kes} for a review or \cite{Montandon:2022ulz} for the specific method we use which is suitable for a relativistic setting.

For the purpose of this discussion, let us work in the continuum limit, taking the number of particles to infinity. Following \cite{Adamek:2015hqa}, we can obtain the particle displacement order by order in perturbation theory by demanding that the Hamiltonian constraint is satisfied on the initial hypersurface. For example, at first order we find
\begin{equation}
    \partial_r^2 \phi_1 + \frac{2}{r} \partial_r \phi_1 - 3 H_\mathrm{in}^2 a_\mathrm{in}^2 \phi_1 = \frac{3}{2} H_\mathrm{in}^2 a_\mathrm{in}^2 \left[3 \phi_1 - \partial_r \delta r_1 - \frac{2}{r} \delta r_1\right]\,,
\end{equation}
where we use that $\psi_1 = \phi_1$ and $\partial_t \phi_1 = 0$ in matter domination. Here, $\delta r_1$ is the displacement from the initial homogeneous particle distribution at first order. The solution for $\delta r_1$ in this case is
\begin{equation}
    \delta r_1 = -\frac{2}{3 H_\mathrm{in}^2 a_\mathrm{in}^2} \partial_r \phi_1 + \frac{5}{r^2} \int_0^r \tilde{r}^2 \phi_1(\tilde{r}) d\tilde{r}\,.
\end{equation}
The boundary condition was chosen $\delta r_1 \rightarrow 0$ as $r \rightarrow 0$, required by regularity. A peculiar feature of this solution is that it does not vanish outside of $r_2$ due to the second term. This term implies that some additional mass is transferred into the curved region from infinity without perturbing the density in the flat region. This can be traced back to an effect called `relativistic mass defect', see \cite{plebanski2024introduction} for a discussion. In short, the active gravitational mass $M(r)$ is not simply the sum of the masses of all particles inside the radius $r$ because gravitational binding energy contributes to $M(r)$. Another way of looking at it is that the initial homogeneous template was prepared for a perfectly flat spatial geometry and, therefore, does not take into account the volume excess inside the curved patch: for a closed-universe simulation, the total proper volume of the initial hypersurface is larger than the proper volume of a flat FLRW model with the same boundary dimensions. If one computes the proper density before any displacement, one finds that it is reduced inside the curved patch compared to the exterior because of the extra volume.

As we need to ensure that the displacement field $\delta r_1$ has a solution compatible with periodic boundary conditions, we add the missing mass to the mass of the particles. Perturbing the particle mass inside the curved region by a constant $\delta m_1 / m$, the Hamiltonian constraint at first order becomes
\begin{equation}
    \partial_r^2 \phi_1 + \frac{2}{r} \partial_r \phi_1 - 3 H_\mathrm{in}^2 a_\mathrm{in}^2 \phi_1 = \frac{3}{2} H_\mathrm{in}^2 a_\mathrm{in}^2 \left[3 \phi_1 - \partial_r \delta r_1 - \frac{2}{r} \delta r_1 + \frac{\delta m_1}{m}\right]\,,
\end{equation}
leading to the solution
\begin{equation}
    \delta r_1 = -\frac{2}{3 H_\mathrm{in}^2 a_\mathrm{in}^2} \partial_r \phi_1 + \frac{r}{3} \frac{\delta m_1}{m} + \frac{5}{r^2} \int_0^r \tilde{r}^2 \phi_1(\tilde{r}) d\tilde{r}\,.
\end{equation}
The boundary condition $\delta r_1 \rightarrow 0$ for $r \rightarrow r_2$ then implies
\begin{equation}
    \frac{\delta m_1}{m} = -\frac{15}{r^3_2} \int_0^{r_2} r^2 \phi_1(r) d r\,.
\end{equation}
At second order, the Hamiltonian constraint gives
\begin{multline}
    \partial_r^2 \phi_2 + \frac{2}{r} \partial_r \phi_2 - 3 H_\mathrm{in} a_\mathrm{in}^2 \partial_t \phi_2 - 3 H_\mathrm{in}^2 a_\mathrm{in}^2 \left(\psi_2 - \phi_1^2\right) + 2 \phi_1 \left(\partial_r^2 \phi_1 + \frac{2}{r} \partial_r \phi_1\right) - \frac{1}{2} \left(\partial_r \phi_1\right)^2 \\= \frac{3}{2} H_\mathrm{in}^2 a_\mathrm{in}^2 \Biggl[- \partial_r \delta r_2 - \frac{2}{r} \delta r_2 + 3 \phi_2 + \frac{\delta m_2}{m}  - \left(3\phi_1 + \frac{\delta m_1}{m}\right) \left(\partial_r \delta r_1 + \frac{2}{r} \delta r_1\right)\qquad\qquad\Biggr.\\\Biggl. + 3 \phi_1 \frac{\delta m_1}{m} + \frac{3}{r^2} \left(\delta r_1\right)^2 + \frac{2}{r} \delta r_1 \partial_r \delta r_1 + \left(\partial_r \delta r_1\right)^2 + \frac{1}{2} v_1^2\Biggr]\,,
\end{multline}
where $v_1$ is the first-order velocity given by $v_1 = -\frac{2}{3} H_\mathrm{in}^{-1} a_\mathrm{in}^{-1} \partial_r \phi_1$.
This equation can be solved for $\delta m_2/m$ in the same way as before.

In practice, instead of adding the extra mass only to the particles inside the curved domain, we evenly distribute it on all particles in the simulation. This implies that when initial conditions are set, indeed some mass needs to be transferred from the exterior to the curved patch, but we have ensured that the displacement field has a solution compatible with boundary conditions. We do not need to derive this solution explicitly, as it will be solved numerically by the iterative method we use to set the initial particle positions.

\subsection{Matter perturbations in the curved patch}
\label{sec:matterperts}

Including matter perturbations in the curved patch can be separated into two problems. First, one has to give a prescription for how to generate initial perturbations in a curved FLRW Universe. For convenience, we will assume that these can be provided in the Poisson gauge on a curved FLRW background. The curved background solution itself is the same in all gauges, so our coordinate transformations \eqref{eq:t-coordtra} and \eqref{eq:r-coordtra} can be used to transform these perturbations into Poisson gauge on a flat FLRW background. We already have everything in our hands for this second step, but the first one is a bit challenging in detail.

As noted in the Introduction, there are several interesting aspects that we want to briefly mention. In a curved Universe one cannot use standard inflation to predict the primordial power spectrum of perturbations, as inflation would always lead to a flat geometry. To the degree to which curvature survives inflation, the initial power spectrum will be sensitive to pre-inflationary physics. We want to emphasise again that this happens at exactly the scale we are interested in: the curvature scale of the Universe also defines the scale at which the power spectrum is modified. Interesting scenarios in which the primordial power spectrum can still be computed self-consistently include open inflation \cite{Linde:1999wv} or the no-boundary state \cite{Halliwell:1984eu} originally proposed by Hartle \& Hawking \cite{Hartle:1983ai}. Having observational evidence for any of these scenarios would be extremely exciting.

Another point to note is that the spectrum of perturbations is usually analysed in terms of eigenfunctions of the Laplace--Beltrami operator on the given background. For a flat background, these would be Fourier modes in a Cartesian basis or the corresponding spherical waves when using spherical polar coordinates. The mode functions in a closed Universe are hyperspherical harmonics, i.e.\ a generalisation of spherical harmonics to the three-sphere, which have a discrete spectrum. Mode functions on the three-hyperboloid can also be constructed and, like in the flat case, they have a continuous spectrum. We refer the interested reader to \cite{Lesgourgues:2013bra} where the implementation of such mode functions in a modern Einstein--Boltzmann solver is discussed in detail. What this means is that the initial perturbations are not easily expressed in terms of Fourier modes in our simulation box. An approximate correspondence can be established on short scales, but on scales where curvature is important, we should expect some corrections.

At present we have not yet developed an efficient method of initialising perturbations on the three-sphere and translating them to the coordinate system of our simulations. The procedure is clear at a conceptual level, but we leave the implementation to future work. In this work, we instead use an approximate method to initialise matter perturbations, ignoring the different nature of mode functions in different geometries. We therefore assume that the mode functions on the scales of interest are well approximated by the Fourier modes of the flat exterior geometry. The only thing we then need to do is to convert comoving wavenumbers between the two different coordinate systems.

In practice, our method of setting initial conditions consists of the following steps. First, we compute a linear prediction for the transfer functions of density, velocity, and gravitational potential for the curved model. This is done with the Einstein--Boltzmann code \texttt{CLASS} \cite{Lesgourgues:2013bra,Blas:2011rf}. Then we convert the wavenumber units from the \texttt{CLASS} output, which are for the curved FLRW model, into the units used in our flat exterior geometry. Unit conversion takes into account the difference in the expansion factor (relevant for comoving units) and the Hubble parameter (relevant because wavenumbers are given in units of $h\,\mathrm{Mpc}^{-1}$). In essence, this rescaling follows the philosophy of the separate universe approach. The resulting transfer functions in the flat coordinates are then used to generate a displacement field and a velocity potential for the particles, as well as the corresponding gravitational potential $\phi$ of the matter perturbations: for each Fourier mode, an initial random value of the primordial curvature perturbation is drawn from a Gaussian whose variance is given by the inflationary power spectrum, and the random value is multiplied by each transfer function to obtain the Fourier modes of the respective perturbation fields. The inflationary power spectrum follows a nearly scale invariant power law parameterised, as usual, with an amplitude $A_\mathrm{s}$ and spectral index $n_\mathrm{s}$, and we are mindful about unit conversion for the pivot scale at which $A_\mathrm{s}$ is defined. We refer the interested reader to Section 6 of \cite{Angulo:2021kes} for the general picture and to Appendix A of \cite{Adamek:2016zes} for details of our implementation. Since we do not want any matter perturbations outside of the curved domain, the final step of our procedure consists of apodising the displacement field, velocity potential, and gravitational field perturbations. This means that we smoothly trim these fields to zero as the distance from the center of the curved domain approaches the inner matching radius $r_1$ from the inside. Given our initial conditions for the simulation without matter perturbations, the perturbations are then linearly added.

\section{Numerical results}
\label{sec:results}

We implement our methodology in the relativistic particle-mesh $N$-body code \texttt{gevolution} \cite{Adamek:2016zes,Adamek:2015eda}. Using a weak-field expansion in the Poisson gauge, this code includes relativistic corrections up to second order in the gravitational potentials $\phi$ and $\psi$, as well as frame dragging to its leading order. It is therefore ideally suited for our purpose. To verify our methodology, we performed three numerical experiments. The first two are designed to validate the approach at the background level. We run an Einstein--de\,Sitter model with large curvature parameter, $\tilde{\Omega}_K = -0.25$, to see if our implementation is robust for a wide range of parameters. We also run a $\Lambda$CDM model with moderate curvature, $\tilde{\Omega}_K = -0.1$, to demonstrate that a cosmological constant can be included and treated correctly. For both simulations, we measure the observed distance--redshift relations for different observers inside the curved domain. In the third experiment, we add matter perturbations in the curved $\Lambda$CDM model and measure observed angular clustering statistics in two redshift bins, again for different observers. This experiment shows that structure formation is consistent with the curved model and that the curved patch satisfies the cosmological principle, that is, that all observers see the same clustering statistics independent of location and direction.

\subsection{Einstein--\texorpdfstring{de\,Sitter}{de Sitter} simulation}

For this test, we run a simulation for an Einstein--de\,Sitter model with $\tilde{\Omega}_K = -0.25$, $\tilde{\Omega}_\mathrm{m} = 1.25$, $\tilde{h} = 0.5$ and no matter perturbations in the curved patch since we want to test the performance of our approach at the level of a curved background cosmology. We use a simulation box of $6000\,h^{-1}\,\mathrm{Mpc}$ that embeds the curved patch inside a region with $r_2 = 2400\,h^{-1}\,\mathrm{Mpc}$ --- note that distance units use the exterior value of $h \approx 0.512$ in our numerical setup. We start the simulation at $\tilde{z}_\mathrm{in} = 25$, which corresponds to $z_\mathrm{in} \approx 25.464$ for our choice of parameters. The simulation has a resolution of $2048^3$ grid points for the mesh and uses $4096^3$ particles.

Using the existing code infrastructure of \texttt{gevolution} to build light cones on the fly, we generate data for the past light cones of two observers. The first one, observer A in Fig.~\ref{fig:setup}, is located at the center of the curved patch and remains at rest with respect to exterior comoving coordinates. The second one, observer B in Fig.~\ref{fig:setup}, is at a distance of about $2250\,h^{-1}\mathrm{Mpc}$ from the center and therefore close to the perimeter. The observed field of view points into the curved patch as illustrated in Fig.~\ref{fig:setup}. As both observers are assumed to be comoving in the curved FLRW background, this observer has a nonvanishing peculiar velocity in the comoving frame of the simulation that, as we remind the reader, is following the expansion of the flat exterior. The peculiar velocity vector of the second observer points straight towards the center of the curved domain, following the flow of matter. We also note that, as discussed earlier, the two observers reach the hypersurface of $\tilde{z} = 0$ at different simulation times (see Fig.~\ref{fig:timeslicings}). This is taken into account when we construct our light cones. For reference, the observation of the central observer is taken at $z = 0$ according to our construction, but the second observer takes the observation at $z \approx 0.01524$. These redshift values correspond to the expansion factor of the simulation box, that is, to some coordinate time of the simulation, not to any observed values of redshifts in the data, $z_\mathrm{obs}$, to which we turn next.

The light cones are post-processed with a ray tracer that solves the geodesic equation and Sachs equations non-perturbatively, following the light rays backward in time from the observation event to the sources, which we take here to be a random sample of simulation particles. We find the unique geodesic to each source using a shooting method; see \cite{Lepori:2020ifz} for more details on our relativistic ray tracer. In this way, we obtain observed positions, redshifts, and angular distances to our sample of matter tracers. These are initially computed in the comoving frame of the simulation. For the second observer, we therefore need to apply a special relativistic boost to obtain the observables in the comoving frame of the matter. We do this by measuring the peculiar velocity at the observer position (here, we use the particle closest to the observer) and applying the appropriate boost transformations to observed redshift (relativistic Doppler boost), position, and angular distance (both are affected by relativistic aberration).

\begin{figure}
  \includegraphics[width=0.49\textwidth]{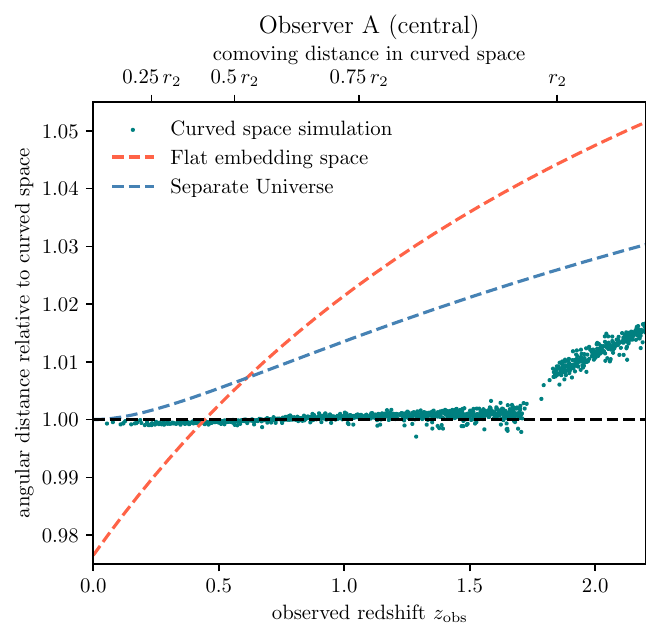}
  \includegraphics[width=0.49\textwidth]{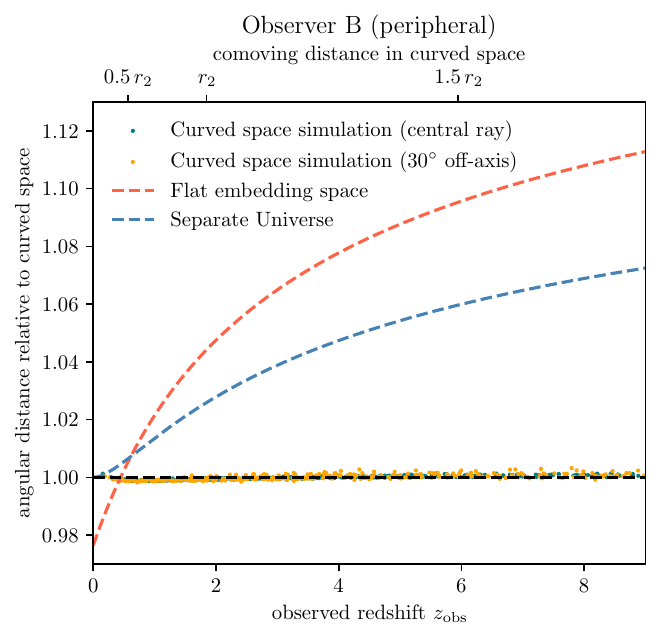}
  \caption{The distance--redshift relations inferred by an observer A at the center of the curved domain (left panel) and by an observer B close to the perimeter when looking at a field of view pointing into the curved domain (right panel). The simulations are consistent with the curved Einstein--de\,Sitter input model ($\tilde{\Omega}_K = -0.25$, $\tilde{\Omega}_\mathrm{m} = 1.25$, $\tilde{h} = 0.5$) to high accuracy for both observers. For the central observer A, all lines of sight are equivalent (up to numerical effects). For observer B, different lines of sight take qualitatively different paths through the curved domain. We therefore compare results for two cases: the line of sight going through the center of the domain, and the lines of sight that are 30 degrees away from that axis. We also indicate the distance--redshift relations in the flat embedding space (red dashed curves) and in the separate universe approximation that does not account for the non-Euclidean spatial geometry (blue dashed curves). The tick labels of the top axis are derived using the distance--redshift relation of the curved space.}
  \label{fig:curvedEdS}
\end{figure}

The final result is independent of any coordinate system used in the computation. In Fig.~\ref{fig:curvedEdS}, we compare the observed angular distance as a function of the observed redshift with the exact relation predicted by the Einstein--de\,Sitter model. The left panel shows results for the observer at the center of the patch, where we find agreement at the level of one part in a thousand up to observed redshift $z_\mathrm{obs} \approx 1.7$, beyond which the past light cone crosses over into the flat embedding domain. This can be seen from the comoving distance in the curved patch, which we indicate in units of $r_2$ on the top horizontal axis. For reference, we also plot the exact distance--redshift relations predicted in the flat exterior model (red dashed curves) and the separate universe approximation (blue dashed curves). The latter consists of using the exact expansion history of the curved model but applying it to spatially flat comoving sections. This approximation leads to a $\sim 1\%$ error at redshift $z_\mathrm{obs} \simeq 1$ at our value of the curvature parameter, originating from the purely geometrical effect of curvature that is independent of the expansion.

The right panel of Fig.~\ref{fig:curvedEdS} shows corresponding results for the second observer, who is located close to the perimeter of the curved domain. Due to this vantage point, the lines of sight for a certain fraction of the sky can extend to much higher redshifts before the past light cone crosses over into the flat embedding domain (see Fig.~\ref{fig:setup} for an illustration). The reduced symmetry of this configuration also means that not all lines of sight take similar paths through the curved patch. To show that the distance--redshift relation remains isotropic from the point of view of the observer as long as the observed sources are contained within the curved patch, we separately plot samples that lie close to the central ray, pointing directly inward, and samples that lie close to $30$ degrees away from that axis. Both samples give measurements consistent with the curved model to one part in a thousand, this time up to redshifts beyond $z_\mathrm{obs} = 8$. At such high redshifts, the separate universe approximation would be off by several percent.

\subsection{Curved \texorpdfstring{$\Lambda$}{Lambda}CDM simulation without matter perturbations}

For this test, we choose somewhat less extreme parameters that are closer to what is required for the forward modelling of real observations. Our curvature parameter, $\tilde{\Omega}_K = -0.1$, is still ruled out by observational constraints, but our aim here is to illustrate the effect of curvature. Importantly, however, this simulation includes a cosmological constant. The cosmological parameters are $\tilde{\Omega}_K = -0.1$, $\tilde{\Omega}_\mathrm{m} = 0.4$, $\tilde{\Omega}_\Lambda \simeq 0.7$, $\tilde{h} = 0.7$, and the remaining energy density is in a small amount of residual radiation (photons and massless neutrinos) which we keep for later convenience but which have no noticeable effect on the background evolution. We use a simulation box of $4500\,h^{-1}\,\mathrm{Mpc}$ that embeds a curved patch inside a region with $r_2 = 1800\,h^{-1}\,\mathrm{Mpc}$. Starting at redshift $\tilde{z}_\mathrm{in} = 15$, we find that the exterior model has $z_\mathrm{in} \approx 15.719$ and $h \approx 0.716$. The simulation has a resolution of $1536^3$ grid points and uses $3072^3$ particles. Note that although these numbers are smaller than for the Einstein--de\,Sitter run, the physical resolution is significantly higher due to the choice of other parameters.

\begin{figure}
  \includegraphics[width=0.49\textwidth]{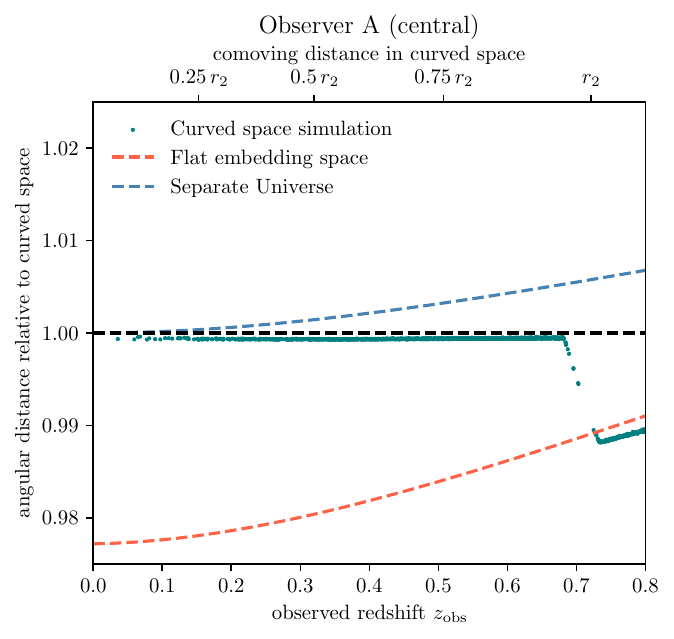}
  \includegraphics[width=0.49\textwidth]{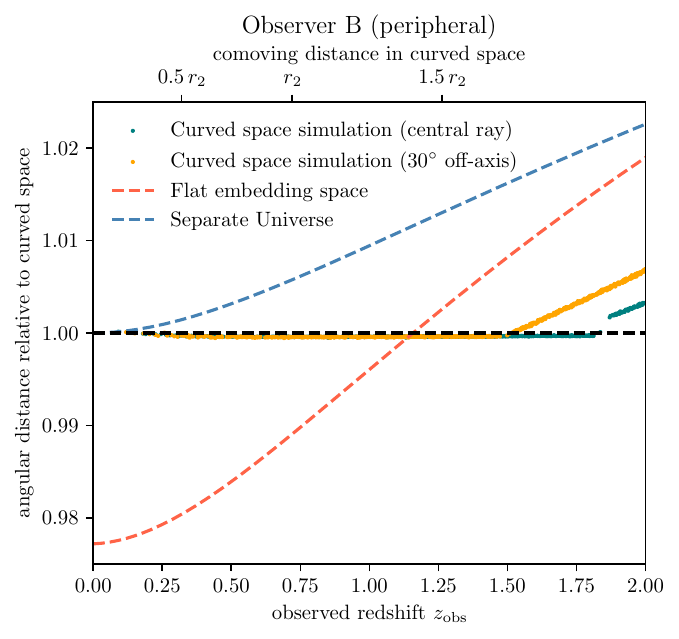}
  \caption{The distance--redshift relation inferred by an observer A at the center of the curved domain (left panel) and by an observer B close to the perimeter when looking at a field of view pointing into the curved domain (right panel). The simulations are consistent with the curved $\Lambda$CDM input model ($\tilde{\Omega}_K = -0.1$, $\tilde{\Omega}_\mathrm{m} = 0.4$, $\tilde{h} = 0.7$) to very high accuracy for both observers. The small residual is a nearly constant offset, corresponding to an error in the fiducial Hubble parameter of about $0.05\,\%$. For the central observer A, all lines of sight are equivalent (up to numerical effects). For observer B, different lines of sight take qualitatively different paths through the curved domain. We therefore compare results for two cases: the line of sight going through the center of the domain, and the lines of sight that are 30 degrees away from that axis. We also indicate the distance--redshift relations in the flat embedding space (red dashed curves) and in the separate universe approximation that does not account for the non-Euclidean spatial geometry (blue dashed curves). The tick labels of the top axis are derived using the distance--redshift relation of the curved space.}
  \label{fig:curvedLCDM}
\end{figure}

We place observer A at the center of the curved patch as before and observer B at a distance of about $1650\,h^{-1}\,\mathrm{Mpc}$ from the center. The coordinate time at which the latter makes the observation corresponds to $z \approx 0.00376$ in the exterior model. Apart from these details, the construction of the observables proceeds along the same steps as outlined for the Einstein--de\,Sitter case. Fig.~\ref{fig:curvedLCDM} shows the results compared to the exact distance--redshift relation predicted in the curved $\Lambda$CDM input cosmology. We find excellent agreement, better than one part in a thousand, for both observers and for the entire range of redshifts for which the respective past light cones are contained within the curved domain. For observer A (left panel of Fig.~\ref{fig:curvedLCDM}), the usable light cone extends to about $z_\mathrm{obs} = 0.6$, whereas observer B (right panel of Fig.~\ref{fig:curvedLCDM}) has access to much higher redshifts. As before, we find that the Hubble diagram remains isotropic within the curved domain. Across all samples, the small residual is a nearly constant offset of about $0.05\%$, which would correspond to an error in the fiducial Hubble parameter of the same order: changing the Hubble parameter of the reference model shifts a curve up or down in these plots without affecting the shape. It is also interesting to note that, despite the smaller curvature parameter of $\tilde{\Omega}_K = -0.1$, the separate universe approximation (blue dashed curves) still exceeds an error of $1\%$ beyond $z_\mathrm{obs} = 1$.

\subsection{Curved \texorpdfstring{$\Lambda$}{Lambda}CDM simulation with matter perturbations}

In our final numerical test, we run the curved $\Lambda$CDM simulation another time, adding matter perturbations in the initial conditions. These are constructed as described in Sec.~\ref{sec:matterperts}, with a primordial scalar amplitude $A_\mathrm{s} = 1.8113 \times 10^{-9}$ at the pivot scale $\tilde{k}_\mathrm{pivot} = 0.05\,\mathrm{Mpc}^{-1}$, where the comoving wavenumber is defined with respect to the curved background, and scalar spectral index $n_\mathrm{s} = 0.9619$. Since the Einstein--Boltzmann solver requires sensible values for the baryon and photon densities, we split $\tilde{\Omega}_\mathrm{m}$ into a baryonic contribution and a CDM contribution, with $\tilde{\Omega}_\mathrm{b} = 0.05$ and $\tilde{\Omega}_\mathrm{cdm} = \tilde{\Omega}_\mathrm{m} - \tilde{\Omega}_\mathrm{b} = 0.35$, respectively. The photon density is fixed by setting the temperature of the cosmic microwave background to $2.7255\,\mathrm{K}$. For completeness, we also include three standard massless neutrino species. The density of these radiation components has a negligible effect on the evolution in the redshift range covered by the simulation, but their presence is of course important for predicting the linear initial conditions of the matter perturbations.

The aim of this test is to measure the observed clustering statistics of matter in the curved patch. In the spirit of our previous test, we want to construct a pure observable that is independent of the coordinate system of the simulation. A suitable observable is given by the angular two-point function of matter within a redshift bin. Its harmonic transform, the angular power spectrum $C_\ell$, can be predicted in linear theory using \texttt{CLASS}. To estimate the angular power spectrum from simulations, we first construct light cones as before, but using a dense sampling of particles. One in every sixteen particles is used as a tracer, arranged in a face-centered cubic packing embedded in the initial particle template. This regular pattern avoids the introduction of shot noise. Note that we keep the observers comoving with the curved background, that is, we use their respective peculiar velocities from the simulation without matter perturbations.

Our observer B uses a cone-shaped field of view covering about 4500 square degrees, so in order to gather more statistics, we place four equivalent copies of this observer inside the curved domain, located at the vertices of an inscribed regular tetrahedron. The idea is that although their fields of view overlap, the data within some thin redshift bins are largely independent and can be treated as separate realisations.

Once the observed positions and redshifts of the matter tracers have been computed, we select two thin redshift bins, centered on $z_\mathrm{obs} = 0.5$ and $z_\mathrm{obs} = 1.0$, using a top-hat selection window of width $\Delta z_\mathrm{obs}=0.05$. Note that the curvature parameter reaches its maximum value of about $12.33\%$ around $\tilde{z}=0.5$, and starts decaying in the later Universe due to the presence of the cosmological constant. We create pixel maps of the particle distribution in these redshift bins using the \texttt{HEALPix} package \cite{Gorski:2004by}, and compute their pseudo-$C_\ell$ estimator with \texttt{NaMaster}\footnote{\href{https://github.com/LSSTDESC/NaMaster}{\faicon{github}~https://github.com/LSSTDESC/NaMaster}} \cite{Alonso:2018jzx}. The redshift bin at $z_\mathrm{obs}=0.5$ is also available on the past light cone of observer A at the center of the curved domain. We want to check if the clustering statistics are independent of observer position, meaning that the cosmological principle is satisfied. We therefore extract four separate fields of view from the central observer's sky, matching the cone-shaped selection of observer B, and apply the same pseudo-$C_\ell$ analysis. To minimise their correlations, the fields of view are again arranged at tetrahedral angles.

\begin{figure}
  \includegraphics[width=\textwidth]{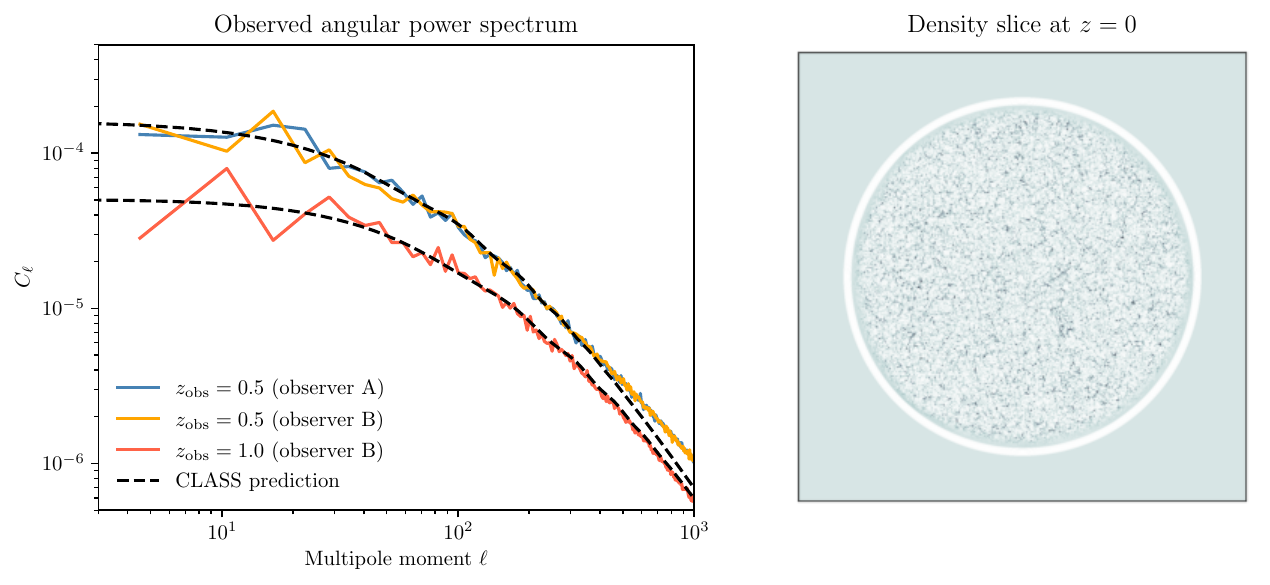}
  \caption{Observed angular power spectra $C_\ell$ in the curved patch once matter perturbations are included (left panel). We show results for two redshift bins, each with a top-hat selection window of width $\Delta z_\mathrm{obs} = 0.05$, centered on $z_\mathrm{obs}=0.5$ and $z_\mathrm{obs}=1.0$. The lower redshift value can be observed from any position within the domain, and we show separate measurements for observer A at the center of the domain and observer B close to the perimeter. For a fair comparison, each observer uses the same survey footprint of about $4500$ square degrees, although observer A has more area at their disposal. To reduce sample variance, we average the measurements over four nearly independent surveys. For observer A this simply means using different patches on the sky, while for observer B we choose four different locations within the curved domain. Our results show good agreement with the linear prediction from \texttt{CLASS} (dashed lines), except for the highest multipoles where linear theory is expected to break down. The right panel shows a density slice of our simulation at redshift $z=0$ for illustration purpose.}
  \label{fig:curvedclustering}
\end{figure}

The left panel of Fig.~\ref{fig:curvedclustering} shows the results of our analysis. The four pseudo-$C_\ell$ estimates for each observer type and each redshift bin are averaged to reduce the sample variance. We confirm that the observed clustering amplitude is consistent with the linear prediction from \texttt{CLASS} (dashed curves) everywhere, except for high multipoles (small scales) at low redshift where we expect an enhancement due to nonlinear structure. We also find that the angular spectra measured by observer A and observer B are statistically indistinguishable, in line with the cosmological principle. For illustration, the right panel of Fig.~\ref{fig:curvedclustering} shows the matter density on a cross-section of the simulation volume at $z = 0$.

Although possible in principle, we did not attempt a comparison to the separate universe approximation for this observable, as this may depend on the way the light cone is constructed. Treating the separate universe model as a flat geometry with a modified expansion history that matches the one of a curved model, one can think of curvature as a `dark fluid' with equation of state $w = -^1\!/\!_3$. The linear prediction in this scenario has been discussed in \cite{DiDio:2016ykq}, where it is directly compared to the linear prediction with geometric curvature. For $\Omega_K = 0.01$, their study indicates a $\sim 1\%$ error on the angular power spectra of matter over a large range of multipoles and redshifts if the geometric effect of curvature is neglected; see their Fig.~2. If this result extrapolates to larger values of curvature, and specific details of their analysis such as redshift binning, linear bias and magnification bias do not change the picture, we might expect $\sim 10\%$ errors on the angular power spectra of matter for the separate universe approximation with our value of $\tilde{\Omega}_K = -0.1$. It is also worth pointing out that the geometric effects of curvature, at least for the examples studied in \cite{DiDio:2016ykq}, turn out to be somewhat larger than the purely dynamical effects that are related to the modification of the expansion history and captured by the separate universe approximation.

\section{Conclusion}
\label{sec:conclusion}

In this paper, we have developed a methodology for running cosmological $N$-body simulations of structure formation in the presence of global spatial curvature. Our relativistic approach takes non-Euclidean distance and volume effects fully into account for the first time. These are mostly relevant for the construction of observables on the light cone that span cosmological distances. By embedding a curved patch of finite size into a periodic embedding space, we can work within the technical constraints of existing $N$-body simulation codes, which commonly rely on periodic boundary conditions. Starting from an exact LTB solution in the synchronous gauge, we present the gauge transformation to Poisson gauge (the gauge used in our relativistic simulation code) to second order, ensuring high accuracy even for large values of curvature. We demonstrate that our approach reproduces some key predictions for the curved models with high accuracy: in the matter-comoving frame, the distance--redshift relation is isotropic around any point in the curved patch and agrees with the one of the curved input model; the angular two-point correlation function of matter in observed reshift bins is also consistent with expectations and satisfies the cosmological principle.

Our work advances the state of the art in forward-modelling observables of models with curvature in several important ways. While the separate universe approximation gives accurate predictions for the clustering statistics on small scales, it lacks a rigorous prescription for constructing large contiguous survey footprints. We have shown that the distance--redshift relation becomes inaccurate in this approximation when taken over large enough distances. This is in line with previous findings that modifying the expansion history alone does not capture the full impact of curvature for observables on our past light cone. In some cases, for example cosmic shear, the effect has been shown to be below sample variance of current surveys for realistic values of curvature \cite{Taylor:2018qda}. Angular clustering statistics, on the other hand, can be more strongly affected \cite{DiDio:2016ykq} and can therefore provide an interesting probe of curvature.

The LTB model has been used previously in simulations of structure formation; see \cite{Alonso:2010zv,Marra:2022ixf} for two relevant examples. Although these studies focus on modelling large spherical inhomogeneities in a standard flat cosmology, it is clear that their methodology could be adapted to our use case. This suggests that Newtonian $N$-body codes should be able to reproduce our results to a very good approximation; see \cite{Marra:2022ixf} for a discussion. However, by not taking the Newtonian limit, our approach is more rigorous, and we clarify several subtleties. For example, we take into account the relativistic mass defect, which is almost $1\%$ in our curved $\Lambda$CDM simulation with $\tilde{\Omega}_K = -0.1$. The size of this effect, like most of the corrections we consider, scales with the difference between Euclidean and curved volume and therefore depends not only on $\tilde{\Omega}_K$ but also on the survey footprint we want to model. The volumes needed to forward-model future surveys may exceed the scale of previous LTB simulations by orders of magnitude, degrading the accuracy of the Newtonian approximation.

While our current implementation only allows for the simulation of a closed universe, the methodology can be generalised to the open case. The minor complication is the need for a compensating mass shell to obtain a plausible sequence of matching radii for the exact LTB solution. We plan to implement this in a future iteration of our code. Furthermore, setting the initial conditions for matter perturbations on scales approaching the curvature scale would require more careful consideration. Our present approach essentially follows the separate universe approximation in assuming that the mode functions on the curved geometry are given by the standard Fourier series instead of hyperspherical harmonics. However, if we were given some initial data on the three-sphere, we would already have computed the coordinate transformation for translating them into our simulation framework. We plan to revisit this aspect in future work.

\acknowledgments

We thank Selim Oueslati for discussions. This work was supported by a grant from the Swiss National Supercomputing Centre (CSCS) under project ID uzh34. We acknowledge financial support from the Swiss National Science Foundation.

\section*{Code availability}

Our implementation is available in the new version 1.3 of \texttt{gevolution}, which can be found at \url{https://github.com/gevolution-code/gevolution-1.3}.

\appendix
\section{Second-order gauge transformation}
\label{app:2nd-order}

Here we present the gauge transformation from the LTB metric \eqref{eq:LTB} in comoving synchronous gauge to Poisson gauge, where it takes the form given in Eq.~\eqref{eq:metric}. Let us expand the metric functions in both gauges around flat FLRW,
\begin{eqnarray}
    \psi &=& \psi_1 + \psi_2\,,\\
    \phi &=& \phi_1 + \phi_2\,,\\
    \frac{\left(\partial_r R\right)^2}{1 + 2 E(r)} &=& a^2(t) \left(1 + 2 y_1 + 2 y_2\right)\,,\\
    R^2 &=& a^2(t) r^2 \left(1 + 2 b_1 + 2 b_2\right)\,.
\end{eqnarray}
Using the expressions for $E(r)$ and $R(t,r)$ from Eqs.~\eqref{eq:E2nd} and \eqref{eq:R2nd}, we find
\begin{eqnarray}
    b_1 &=& \frac{1}{3} \left(1-\frac{a(t)}{a_\mathrm{in}}\right) f(r)\,,\\
    y_1 &=& b_1 + r \partial_r b_1 + r h'(r)\,,\\
    b_2 &=& -\frac{1}{6} b_1 \left(1+\frac{1}{7}\frac{a(t)}{a_\mathrm{in}}\right) f(r)\,,\\
    y_2 &=& \left(2 y_1 - \frac{1}{7} f(r)\right) r h'(r) + \frac{1}{2}\left(b_1 + r \partial_r b_1\right)^2-\left(b_1 + 2 r \partial_r b_1\right)\left(\frac{1}{3} - \frac{1}{7}\frac{a(t)}{a_\mathrm{in}}\right) f(r)\,,\qquad
\end{eqnarray}
where we defined the shorthand
\begin{equation}
    h(r) = \frac{5}{6} H_\mathrm{in}^2 a_\mathrm{in}^2 \int^r_{r_2} \tilde{r} f(\tilde{r}) d\tilde{r}\,.
\end{equation}

Since the line element $ds^2$ is invariant, we have
\begin{equation}
\label{eq:metrictransform}
    g_{\mu\nu}^\mathrm{(P)} = g_{\alpha\beta}^\mathrm{(syn)} \frac{dx^\alpha_\mathrm{(syn)}}{dx^\mu_\mathrm{(P)}}  \frac{dx^\beta_\mathrm{(syn)}}{dx^\nu_\mathrm{(P)}}\,,
\end{equation}
where both metrics have to be evaluated at the same space-time event (which has different coordinate values in the two coordinate systems). The first order of Eq.~\eqref{eq:metrictransform} gives the following relations,
\begin{eqnarray}
    \psi_1 - \partial_t T_1 &=& 0\,, \qquad \text{(}t\text{-}t\text{-component)}\\
    a^2 \partial_t L_1 - \partial_r T_1&=& 0\,, \qquad \text{(}t\text{-}r\text{-component)}\label{eq:1st-tr}\\
    \phi_1 + y_1 + H T_1 + \partial_r L_1 &=& 0\,, \qquad \text{(}r\text{-}r\text{-component)}\\
    \phi_1 + b_1 + H T_1 + \frac{1}{r} L_1 &=& 0\,. \qquad \text{(}\theta\text{-}\theta\text{-component)}
\end{eqnarray}
Combining the latter two, we can solve for $L_1$,
\begin{equation}
    r \partial_r \left(\frac{L_1}{r}\right) = - r \partial_r b_1 - r h'(r) \qquad \Rightarrow \qquad L_1 = -r b_1 - r h(r)\,.
\end{equation}
Inserting this solution into Eq.~\eqref{eq:1st-tr} fixes $T_1$,
\begin{equation}
    \frac{H a^3}{3 a_\mathrm{in}} r f(r) = \partial_r T_1 \qquad \Rightarrow \qquad T_1 = \frac{2}{5 H} h(r)\,,
\end{equation}
where we use the fact that $H^2 a^3$ is independent of time. We can finally read off the solutions for $\phi_1$ and $\psi_1$,
\begin{equation}
    \psi_1 = \phi_1 = \frac{3}{5} h(r)\,.
\end{equation}

The second order of Eq.~\eqref{eq:metrictransform} gives the following relations,
\begin{eqnarray}
    \psi_2 - \partial_t T_2 &=& \frac{1}{2} \left(\partial_t T_1\right)^2 - \frac{1}{2} a^2 \left(\partial_t L_1\right)^2 - \psi_1^2\,, \label{eq:2nd-tt}\\
    a^2 \partial_t L_2 - \partial_r T_2&=& \partial_t T_1 \partial_r T_1 - a^2 \partial_t L_1\left(2 y_1 + \partial_r L_1 + 2 H T_1\right)\,, \label{eq:2nd-tr}\\
    \phi_2 + y_2 + H T_2 + \partial_r L_2 &=& \frac{1}{2a^2} \left(\partial_r T_1\right)^2 - T_1 \left(2 H y_1 + \partial_t y_1 +\frac{1}{4} H^2 T_1\right) - L_1 \partial_r y_1 \nonumber\\&&- \partial_r L_1 \left(2 y_1 + 2 H T_1 + \frac{1}{2} \partial_r L_1\right) + \phi_1^2\,, \\
    \phi_2 + b_2 + H T_2 + \frac{1}{r} L_2 &=& - T_1 \left(2 H b_1 + \partial_t b_1 +\frac{1}{4} H^2 T_1\right) - L_1 \partial_r b_1\nonumber\\&&- \frac{1}{r} L_1 \left(2 b_1 + 2 H T_1 + \frac{1}{2 r} L_1\right) + \phi_1^2\,. \label{eq:2nd-qq}
\end{eqnarray}
Subtracting the two last equations gives
\begin{multline}
    r \partial_r \left(\frac{L_2}{r}\right) = b_2 - y_2 + \frac{1}{2a^2} \left(\partial_r T_1\right)^2 - T_1 \Bigl(2 H \left(y_1 - b_1\right) + \partial_t\left(y_1 - b_1\right)\Bigr) - L_1 \partial_r \left(y_1 - b_1\right)\\-2 H T_1 r \partial_r \left(\frac{L_1}{r}\right) - 2 y_1 \partial_r L_1 + \frac{2}{r} b_1 L_1 - \frac{1}{2} \left(\partial_r L_1\right)^2 + \frac{1}{2} \left(\frac{L_1}{r}\right)^2\,.
\end{multline}
Using the first-order solutions and after some algebra, this simplifies to
\begin{multline}
    \partial_r \left(\frac{L_2}{r}\right) = \frac{1}{2} \partial^2_r \left[r\left(b_1 + h\right)^2\right] + \frac{1}{2} \partial_r \left[b_1 \left(f - r h'\right)\right] - \frac{13}{42} \frac{a}{a_\mathrm{in}} \partial_r \left(b_1 f\right) + \frac{2}{15} \frac{a}{a_\mathrm{in}} \left(f h\right)' \\
    + \left(\frac{10}{21} - \frac{2}{5} \frac{a}{a_\mathrm{in}}\right) f h' - \frac{3}{2} r \left(h'\right)^2\,,
\end{multline}
and hence
\begin{multline}
    L_2 = \frac{1}{2} r \partial_r \left[r\left(b_1 + h\right)^2\right] + \frac{1}{2} r b_1 \left(f - r h'\right) - \frac{13}{42} \frac{a}{a_\mathrm{in}} r b_1 f + \frac{2}{15} \frac{a}{a_\mathrm{in}} r f h \\+ r \int_{r_2}^r h'(\tilde{r}) \left[\left(\frac{10}{21} - \frac{2}{5} \frac{a}{a_\mathrm{in}}\right) f(\tilde{r}) - \frac{3}{2} \tilde{r} h'(\tilde{r})\right] d\tilde{r}\,.
\end{multline}

We may now compute $a^2 \partial_t L_2$ that appears in Eq.~\eqref{eq:2nd-tr}. Using the definition of $h(r)$ together with the fact that $H^2 a^3$ is independent of time, we can derive
\begin{equation}
    a^2 \partial_t L_2 = \frac{1}{5 H} \left[\frac{4}{7} \frac{a}{a_\mathrm{in}} f h' -4 r b_1 h'' - \left(\frac{20}{21} f - \frac{4}{5} h + r h'\right) h' - 2 r h h'' - \frac{12}{5} r \int_{r_2}^r \left(h'(\tilde{r})\right)^2 \frac{d\tilde{r}}{\tilde{r}}\right]\,.
\end{equation}
We therefore find
\begin{equation}
    \partial_r T_2 = \frac{1}{5 H} \left[\left(\frac{4}{7} \frac{a}{a_\mathrm{in}} f - \frac{20}{21} f + r h'\right) h' -2 r \left(b_1 + h\right) h'' - \frac{2}{5} \partial_r \left(h^2\right) - \frac{12}{5} r \int_{r_2}^r \left(h'(\tilde{r})\right)^2 \frac{d\tilde{r}}{\tilde{r}}\right]\,,
\end{equation}
and hence
\begin{multline}
    T_2 = \frac{3 h^2}{25 H} -\frac{2 r h h'}{5 H} - \frac{r b_1 h'}{5 H} -\frac{2}{5 H} \int_{r_2}^r h'(\tilde{r})\left[\left(\frac{10}{21}-\frac{2}{7} \frac{a}{a_\mathrm{in}}\right) f(\tilde{r}) - \frac{21}{10} \tilde{r} h'(\tilde{r}) \right] d\tilde{r} \\- \frac{6}{25 H} r^2 \int_{r_2}^r \left(h'(\tilde{r})\right)^2 \frac{d\tilde{r}}{\tilde{r}}\,,
\end{multline}
where we use integration by parts on some of the terms to simplify the expression.

Next, we compute $\partial_t T_2$, as it enters Eq.~\eqref{eq:2nd-tt},
\begin{equation}
    \partial_t T_2 = \frac{3}{2} H T_2 + \frac{1}{15} \frac{a}{a_\mathrm{in}} r f h' + \frac{4}{35} \frac{a}{a_\mathrm{in}} \int_{r_2}^r h'(\tilde{r}) f(\tilde{r}) d\tilde{r}\,,
\end{equation}
and therefore
\begin{equation}
    \psi_2 = -\frac{3}{5}\left[r h h' + \frac{1}{2} r b_1 h' + \int_{r_2}^r h'(\tilde{r})\left(\frac{10}{7} b_1(t,\tilde{r}) - \frac{21}{10} \tilde{r} h'(\tilde{r}) \right) d\tilde{r} + \frac{3}{5} r^2 \int_{r_2}^r \left(h'(\tilde{r})\right)^2 \frac{d\tilde{r}}{\tilde{r}}\right]\,.
\end{equation}

Lastly, we solve for $\phi_2$, e.g.\ using Eq.~\eqref{eq:2nd-qq},
\begin{equation}
    \phi_2 = -\frac{3}{5}\left[r h h' + \frac{1}{2} r b_1 h' + \int_{r_2}^r h'(\tilde{r})\left(\frac{10}{7} b_1(t,\tilde{r}) - \frac{11}{10} \tilde{r} h'(\tilde{r}) \right) d\tilde{r} - \frac{2}{5} r^2 \int_{r_2}^r \left(h'(\tilde{r})\right)^2 \frac{d\tilde{r}}{\tilde{r}} \right]\,.
\end{equation}

\section{An exact solution in the Schwarzschild region}
\label{app:Schwarzschild}

Outside of the curved patch, i.e.\ in the vacuum region and the flat FLRW embedding spacetime, the model corresponds exactly to the one of Einstein \& Straus \cite{Einstein:1945id}. It describes a Schwarzschild spacetime embedded in an expanding FLRW solution, with appropriate junction conditions satisfied at the interface. Here we present the metric of this entire manifold, valid on both sides of the junction hypersurface, $\left\{r = r_2\right\}$, written in terms of FLRW comoving coordinates and using the Poisson gauge. Specifically, our metric takes the form given in Eq.~\eqref{eq:metric}, with $a(t) \propto t^{2/3}$ the FLRW scale factor and $\phi = \psi = 0$ in the FLRW region. Our aim here is to obtain exact expressions for $\phi$ and $\psi$ in the vacuum (Schwarzschild) region.

Instead of attempting to derive the gauge transformation from the LTB metric in a non-perturbative way, we may instead consider solving Einstein's equations directly, given that the right-hand side vanishes in vacuum. Due to symmetries, these reduce to only three coupled partial differential equations,
\begin{eqnarray}
    e^{2\phi} \left[\partial_r^2 \phi + \frac{2}{r} \partial_r \phi - \frac{1}{2} \left(\partial_r \phi\right)^2\right] + \frac{3}{2} a^2 \left(\partial_t \phi - H\right)^2 e^{-2\psi} &=&0\,,\label{eq:ESHamiltonianconstraint}\\
    \partial_t \partial_r \phi - \left(\partial_t \phi - H\right) \partial_r \psi &=& 0\,,\label{eq:ESmomentumconstraint}\\
    \partial_r^2\left(\phi - \psi\right) - \frac{1}{r} \partial_r\left(\phi - \psi\right) + \left(\partial_r \phi - \partial_r \psi\right)^2 - 2 \left(\partial_r \psi\right)^2 &=& 0\,.
\end{eqnarray}
The momentum constraint \eqref{eq:ESmomentumconstraint} implies
\begin{equation}
    \partial_r \ln \left(\partial_t \phi - H\right) = \partial_r \psi\,,
\end{equation}
which we may integrate on a constant-time hypersurface to get
\begin{equation}
\label{eq:AppB5}
    \partial_t \phi - H = -H e^\psi\,.
\end{equation}
The constant of integration is fixed by the requirement that both $\psi$ and $\partial_t \phi$ must vanish identically at the junction. We can insert this solution into the Hamiltonian constraint \eqref{eq:ESHamiltonianconstraint} to get a nonlinear ordinary differential equation (ODE) for $\phi$ on a constant-time hypersurface,
\begin{equation}
    \partial_r^2 \phi + \frac{2}{r} \partial_r \phi - \frac{1}{2} \left(\partial_r \phi\right)^2 + \frac{3}{2} a^2 H^2 e^{-2\phi} = 0\,.
\end{equation}
To get further analytical insight, we make the following change of variables,
\begin{eqnarray}
    u = \sqrt{3} a H r e^{-\phi} &\quad\Rightarrow\quad& \phi = \ln\left(\sqrt{3} a H r_2\right) + \ln\frac{r}{r_2} - \ln u\,,\\
    x = -\ln\frac{r}{r_2} &\quad\Rightarrow\quad& r = r_2 e^{-x}\,,
\end{eqnarray}
which casts the ODE into the convenient form
\begin{equation}
    -\frac{2 u''}{u} + \left(\frac{u'}{u}\right)^2 + u^2 + 1 = 0\,,
\end{equation}
where a prime denotes the derivative with respect to the argument, $x$ in this case. Multiplying with $u'$ yields
\begin{equation}
    \left(-\frac{\left(u'\right)^2}{u} + \frac{1}{3} u^3 + u\right)' = 0\,,
\end{equation}
which integrates to
\begin{equation}
    \left(u'\right)^2 = \frac{1}{3} u^4 + u^2 + C u\,.
\end{equation}
We can fix the constant of integration by noting that at $x = 0$ ($r = r_2$) we have $u(0) = \sqrt{3} a H r_2 \equiv u_0$ and $u'(0) = -u_0$ because both $\phi$ and $\partial_r \phi$ must vanish. We find $C = -\frac{1}{3} u_0^3$, and therefore
\begin{equation}
    x(u) = -\int_{u_0}^u \frac{d\tilde{u}}{\sqrt{\frac{1}{3}\tilde{u}^4 + \tilde{u}^2 - \frac{1}{3}u_0^3\tilde{u}}}\,.
\end{equation}

\begin{figure}
  \includegraphics[width=0.49\textwidth]{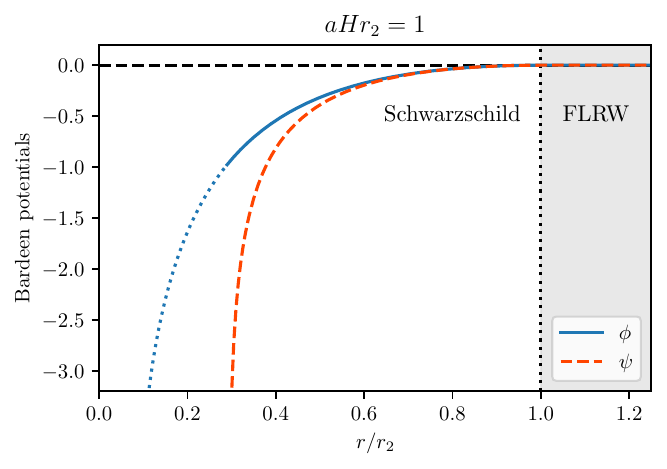}
  \includegraphics[width=0.49\textwidth]{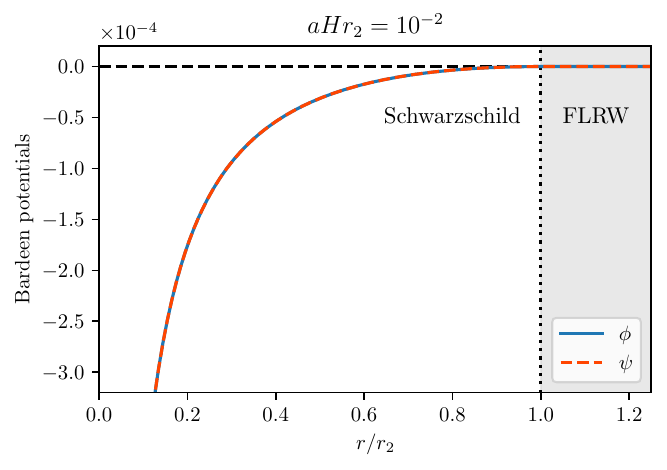}
  \caption{Exact solutions for the two Bardeen potentials, $\phi$ and $\psi$, in the Einstein--Straus model written in Poisson gauge adapted to the FLRW embedding spacetime. The left panel shows the two potentials at early time when the Schwarzschild radius is comparable to $a r_2$, the proper radius of the vacuum region. Both potentials diverge for small finite radii, with $\psi$ diverging at the larger of the two. As the coordinate system becomes invalid at that point, we indicate the continuation of $\phi$ as dotted curve. The right panel shows the situation at a later time when the proper radius of the vacuum region has grown by a factor of $10^4$ with respect to the Schwarzschild radius. Note that $a H r_2 = \sqrt{2 G M} / \sqrt{a r_2}$. At fixed comoving radius far away from the event horizon of the black hole the potentials evolve as $\sim a^2 H^2 \propto a^{-1}$. This behaviour is consistent with a point-mass potential evaluated in comoving coordinates. Note, however, that the potentials and their gradients go to zero at $r \rightarrow r_2$ and remain exactly zero for larger radii.}
  \label{fig:ESpotentials}
\end{figure}

According to a result by Biermann \& Weierstrass (see Appendix A of \cite{Cieslik:2022uki} for details, including original references), $u$ can be expressed in terms of the Weierstrass elliptic function $\wp(-x; g_2, g_3)$,
\begin{equation}
    u = u_0 + \frac{\frac{1}{3} u_0^3 + \frac{1}{2} u_0 \left(u_0^2 + 2\right)\left[\wp(-x) -\frac{1}{6} u_0^2 - \frac{1}{12}\right] - u_0 \wp'(-x)}{2 \left[\wp(-x) -\frac{1}{6} u_0^2 - \frac{1}{12}\right]^2 - \frac{1}{6} u_0^2}\,,
\end{equation}
with the Weierstrass invariants $g_2 = \frac{1}{12}$ and $g_3 = -\frac{1}{432} (2 + u_0^6)$.

After changing variables back from $x$ to $r$, an exact expression for $\phi$ is given by
\begin{equation}
    \phi(t, r) = \ln\frac{r}{r_2} - \ln\frac{u(t,r)}{u_0(t)}\,,
\end{equation}
where we note that the time dependence is encoded entirely in $u_0$, which also enters the definition of $g_3$. We remind the reader that the external FLRW solution is given by $a \propto t^{2/3}$ and hence $H = \frac{2}{3} t^{-1}$. An expression for $\psi$ can now be derived from Eq.~\eqref{eq:AppB5}. The solutions for $\phi$ and $\psi$ are illustrated in Fig.~\ref{fig:ESpotentials}.

To our knowledge, this is the first time that this exact form of the Schwarzschild metric has been reported.

\bibliographystyle{JHEP}
\bibliography{biblio}
\end{document}